\documentclass[12pt]{iopart}
\usepackage[dvips]{graphicx}

\font\bsy=cmbsy12
\def\bfnabla{\vec{\hbox{\bsy\char114}}}

\font\bsy=cmbsy12
\def\enabla{\hbox{\bsy\char114}}

\begin{document}

\title[Correlations and sum rules in a half-space quantum plasma]{Correlations 
and sum rules in a half-space for a quantum two-dimensional one-component 
plasma}
 
\author{B. Jancovici\dag\ and L. {\v S}amaj\dag\ddag}
\address{\dag\ Laboratoire de Physique Th\'eorique, Universit\'e
Paris-Sud (Unit\'e Mixte de Recherche no. 8627 - CNRS),
91405 Orsay, France}
\address{\ddag\ Institute of Physics, Slovak Academy of Sciences,
D\'ubravsk\'a cesta 9, 845 11 Bratislava, Slovakia}

\eads{\mailto{Bernard.Jancovici@th.u-psud.fr},
\mailto{Ladislav.Samaj@savba.sk}}

\begin{abstract}
This paper is the continuation of a previous one [L. {\v S}amaj and
B. Jancovici, 2007 {\it J. Stat. Mech.} P02002]; for a nearly classical
quantum fluid in a half-space bounded by a plain plane hard wall (no image
forces), we had generalized the Wigner-Kirkwood expansion of the equilibrium
statistical quantities in powers of Planck's constant $\hbar$. As a model
system for a more detailed study, we consider the quantum two-dimensional
one-component plasma: a system of charged particles of one species, 
interacting through the logarithmic Coulomb potential in two dimensions,
in a uniformly charged background of opposite sign, such that the total charge
vanishes. The corresponding classical system is exactly solvable in a variety
of geometries, including the present one of a half-plane, when $\beta e^2=2$, 
where $\beta$ is the inverse temperature and $e$ is the charge of a particle:
all the classical $n$-body densities are known. In the present paper, we have
calculated the expansions of the quantum density profile and truncated
two-body density up to order $\hbar^2$ (instead of only to order $\hbar$ in
the previous paper). These expansions involve the classical $n$-body densities
up to $n=4$, thus we obtain exact expressions for these quantum expansions in
this special case. 

For the quantum one-component plasma, two sum rules involving the truncated
two-body density (and, for one of them, the density profile) have been
derived, a long time ago, by heuristic macroscopic arguments: one sum rule is
about the asymptotic form along the wall of the truncated two-body density,
the other one is about the dipole moment of the structure factor. In the
two-dimensional case at $\beta e^2=2$, we have now explicit expressions up to
order $\hbar^2$ of these two quantum densities, thus we can microscopically
check the sum rules at this order. The checks are positive, reinforcing the
idea that the sum rules are correct.  
\end{abstract}


\medskip


\maketitle

\eqnobysec 

\section{Introduction}
The model under consideration is the two-dimensional (2D) one-component plasma 
(also called jellium). This model consists of one species of charged particles 
in a plane. Each particle has a charge $e$ and a mass $m$. Two particles, at a 
distance $r$ from each other, interact through the 2D Coulomb 
interaction $v(r)$. This interaction is determined by the 2D 
Poisson equation $\enabla^2 v(r)=-2\pi e^2\delta(\mathbf{r})$, the solution 
of which is $v(r)=-e^2\ln(r/r_0)$, where $r_0$ is an arbitrary length which
only fixes the zero of this potential. In addition, there is a charged uniform 
background of charge density opposite to the particle charge, so that the total
system is neutral.  

When the inverse temperature $\beta$ is such that the dimensionless
coupling constant $\beta e^2=2$, the equilibrium statistical mechanics of the 
classical (i.e. non-quantum) system is completely solvable
in a variety of geometries, in particular when the background and particles
are confined into a half-space by an impenetrable rectilinear plain hard wall
(there are no image forces): all the classical $n$-body densities are known
\cite{Janco,CJB}.

In its three-dimensional version, the quantum one-component plasma is not only
of academic interest. It has been used, in first approximation, as a model for
the electrons of a metal \cite{NP}. In the present geometry of a half-space,
this model might be used for describing what happens near the surface of the
metal. Since the logarithmic interaction $v(r)$ is the Coulomb potential in
two dimensions, the 2D one-component plasma is expected to have
general features which mimic those of the 3D one. 

Under certain conditions, the equilibrium properties of an infinite quantum
fluid, in the nearly classical regime, can be expanded in powers of Planck's
constant $\hbar$: this is the Wigner-Kirkwood expansion \cite{Wigner,Kirkwood}.
In a previous paper \cite{WK}, we have generalized the Wigner-Kirkwood 
expansion to the case of a quantum fluid occupying a half space. 
We have obtained expressions for the first quantum correction of order 
$\hbar$ to the density profile and to the two-body density, in terms of some 
$n$-body densities of the classical fluid. 
In section 3 of the present paper we extend these calculations to order 
$\hbar^2$. 
Actually, instead of $\hbar$, we use the thermal de Broglie wavelength 
proportional to it, $\lambda=\hbar\sqrt{\beta/m}$.

For the quantum one-component plasma in a half space, by heuristic 
macroscopic methods, some sum rules involving the one-body density 
profile and the two-body density have been derived. 
The purpose of the present paper is to check these sum rules at order 
$\hbar^2$, in the special case of the 2D one-component plasma 
at $\beta e^2=2$, using the generalized Wigner-Kirkwood expansion, 
which is a microscopic approach.
 
The paper is organized as follows. 
Section 2 brings a recapitulation of the method for constructing
the expansion of the quantum Boltzmann density in configuration\linebreak
$\vec{\bf r}$-space for fluids constrained to a half-space \cite{WK}.
This expansion is subsequently used in section 3 to compute the quantum 
one-body profile and two-body density to order $\hbar^2$ for 
the studied 2D one-component plasma.
In section 4, three sum rules are reviewed: a perfect screening rule, 
a sum rule about the asymptotic form of the two-body density along the wall, 
and a dipole sum rule; their expansions to order $\hbar^2$ are given. 
In section 5, some classical $n$-body correlation functions, 
which are needed, are studied. 
In section 6, the perfect screening sum rule is shown to hold, at order 
$\hbar^2$, for the quantum 2D one-component plasma at $\beta=2$. 
In sections 7 and 8, the same is done for the asymptotic form and 
the dipole sum rules, respectively. 
Section 9 is a Conclusion.

\section{Boltzmann density for the half-space geometry}
We first consider a general quantum system of $N$ identical particles 
$j=1,2,\ldots,N$ of mass $m$, formulated in $\nu$ space dimensions.
Particle position vectors ${\bf r}_1,{\bf r}_2,\ldots,{\bf r}_N$ 
are confined to the half-space $\Lambda$ defined by Cartesian
coordinates ${\bf r}=(x>0,{\bf r}^{\perp})$, where 
${\bf r}^{\perp}\in {\rm R}^{\nu-1}$ denotes the set of
$(\nu-1)$ unbounded coordinates normal to $x$.
As usual, we start with a finite $N$ and a finite volume $|\Lambda|$, and later
we take the thermodynamic limit $N$ and $|\Lambda|$ going to infinity (when
$\Lambda$ becomes a half space) with a finite mean number density
$n=N/|\Lambda|$. 
The hard wall in the complementary half-space $\bar{\Lambda}$ of points
${\bf r} = (x<0,{\bf r}^{\perp})$ is considered to be impenetrable
to particles, i.e. the wavefunctions of the particle system vanish
as soon as one of the particles lies at the wall. 
For the sake of brevity, we denote the $\nu N$-dimensional position
vector in configuration space by 
$\vec{\bf r}=({\bf r}_1,{\bf r}_2,\ldots,{\bf r}_N)$ and the corresponding
gradient by $\bfnabla = (\enabla_1,\enabla_2,\ldots,\enabla_N)$.
In the absence of a magnetic field, the Hamiltonian of the particle
system is given by
\begin{equation} \label{2.1}
H = \frac{1}{2 m} \left( - {\rm i} \hbar \bfnabla \right)^2
+ V(\vec{\bf r}) ,
\end{equation}
where $\hbar$ stands for Planck's constant and $V(\vec{\bf r})$ 
is the total interaction potential.

For infinite (bulk) quantum fluids of particles interacting via pairwise
sufficiently smooth interactions with neglected fermion/boson exchange
effects, Wigner \cite{Wigner} and Kirkwood \cite{Kirkwood} constructed 
a semiclassical expansion of the Boltzmann density in configuration space 
(at inverse temperature $\beta$),
$B_{\beta}(\vec{\bf r}) = 
\langle \vec{\bf r}\vert {\rm e}^{-\beta H} \vert \vec{\bf r} \rangle$,
in even powers of the thermal de Broglie wavelength 
$\lambda = \hbar (\beta/m)^{1/2}$. 
Recently \cite{WK}, we have generalized the Wigner-Kirkwood method
to quantum fluids constrained to the above defined half-space $\Lambda$.
The final result for the Boltzmann density in configuration space was
obtained as a series
\begin{equation} \label{2.2}
B_{\beta}(\vec{\bf r}) = \sum_{n=0}^{\infty} B_{\beta}^{(n)}(\vec{\bf r}) ,
\end{equation}
where the terms $B_{\beta}^{(n)}(\vec{\bf r})$ with $n=0,1,2,\ldots$
can be calculated systematically with the aid of an operator technique.

The result for $B_{\beta}^{(0)}(\vec{\bf r})$ was found in the form
\begin{equation} \label{2.3} 
B_{\beta}^{(0)}(\vec{\bf r}) = \frac{1}{(\sqrt{2\pi} \lambda)^{\nu N}} 
{\rm e}^{-\beta V} \prod_{j=1}^N \left( 1 - {\rm e}^{-2x_j^2/\lambda^2} 
\right) .  
\end{equation}
The bulk counterpart of this term corresponds to the classical 
Boltzmann density $\propto {\rm e}^{-\beta V}$. 
Here, each particle gets an additional ``boundary'' factor 
$1-\exp(-2 x^2/\lambda^2)$ which goes from $0$ at the $x=0$ boundary
to $1$ in the bulk interior $x\to\infty$ on the length scale $\sim\lambda$.
The product of boundary factors then ensures that the quantum Boltzmann
density vanishes as soon as one of the particles lies on the boundary.
The dependence of the boundary factor on the de Broglie wavelength
$\lambda$ is non-analytic; this fact prevents one from a simple classification
of contributions to the Boltzmann density according to integer powers 
of $\lambda$ like it is in the bulk case.
However, when in the calculation of statistical averages the exponential part 
$\exp(-2 x^2/\lambda^2)$ of the boundary factor is integrated over 
the $x$-coordinate, the analyticity of the result in the parameter $\lambda$ 
is restored.
At this stage we only notice that when the product of boundary 
factors is expanded as follows
\begin{equation} \label{2.4}
\prod_{j=1}^N \left( 1 - {\rm e}^{-2x_j^2/\lambda^2} \right)
= 1 - \sum_{j=1}^N {\rm e}^{-2x_j^2/\lambda^2} 
+ \frac{1}{2!} \sum_{j,k=1\atop (j\ne k)}^N {\rm e}^{-2x_j^2/\lambda^2} 
{\rm e}^{-2x_k^2/\lambda^2} + \cdots , 
\end{equation} 
the integration of each exponential term $\exp(-2 x^2/\lambda^2)$
over $x$ produces one $\lambda$-factor as the result of 
the substitution of variables $x=\lambda x'$.
 
The result for $B_{\beta}^{(1)}(\vec{\bf r})$ reads
\begin{eqnarray}
B_{\beta}^{(1)}(\vec{\bf r}) & = & 
\frac{1}{(\sqrt{2\pi} \lambda)^{\nu N}} \Bigg\{ 
\sum_{k=1}^N \prod_{j=1\atop (j\ne k)}^N 
\left( 1 - {\rm e}^{-2x_j^2/\lambda^2} \right) 
x_k {\rm e}^{-2x_k^2/\lambda^2}
\frac{\partial }{\partial x_k} {\rm e}^{-\beta V} \nonumber \\ 
& & + {\rm e}^{-\beta V} \prod_{j=1}^N 
\left( 1 - {\rm e}^{-2x_j^2/\lambda^2} \right)
\lambda^2 \left[ - \frac{\beta}{4} \bfnabla^2 V +
\frac{\beta^2}{6} \left( \bfnabla V \right)^2 \right] \Bigg\} . \label{2.5}
\end{eqnarray}
Here, the dependence on $\lambda$ appears also via
the combination $x \exp(-2x^2/\lambda^2)$.
This function has a maximum of order $\lambda$ and therefore 
it is a legitimate expansion parameter.
When integrated over the particle coordinate $x$, it gives a contribution 
of order $\lambda^2$, ``weaker'' than $\lambda$.

Keeping all contributions up to the order $\lambda^2$ in the Boltzmann
term $B_{\beta}^{(2)}(\vec{\bf r})$, one has  
\begin{equation} \label{2.6}
\fl B_{\beta}^{(2)}(\vec{\bf r}) = 
\frac{1}{(\sqrt{2\pi} \lambda)^{\nu N}} {\rm e}^{-\beta V} 
\prod_{j=1}^N \left( 1 - {\rm e}^{-2x_j^2/\lambda^2} \right)
\lambda^2 \left[ \frac{\beta}{6} \bfnabla^2 V 
- \frac{\beta^2}{8} \left( \bfnabla V\right)^2 \right] 
+ o(\lambda^2) . 
\end{equation}
Note that, with regard to the equality
\begin{equation} \label{2.7}
\bfnabla^2 {\rm e}^{-\beta V} = {\rm e}^{-\beta V} \left[
\beta^2 \left( \bfnabla V \right)^2 - \beta \bfnabla^2 V \right],
\end{equation}
one can eliminate the squared gradient term in favour of the Laplacian term
in equations (\ref{2.5}) and (\ref{2.6}).

The first three Boltzmann terms (\ref{2.3}), (\ref{2.5}) and (\ref{2.6})
exhibit properties analogous to their bulk counterparts:
the maximum of $B_{\beta}^{(n)}(\vec{\bf r})$ is of order $\lambda^n$.
We anticipate that this formal structure is also maintained on higher levels.
As a consequence, the knowledge of the first three Boltzmann terms 
$B_{\beta}^{(0)}$, $B_{\beta}^{(1)}$ and $B_{\beta}^{(2)}$    
is sufficient in order to obtain the expansion of the quantum Boltzmann
density up to the $\lambda^2$ order:
\begin{equation} \label{2.8}
B_{\beta}(\vec{\bf r}) = B_{\beta}^{(0)}(\vec{\bf r}) +
B_{\beta}^{(1)}(\vec{\bf r}) + B_{\beta}^{(2)}(\vec{\bf r}) + o(\lambda^2) .
\end{equation}

The quantum fluid of present interest is the one-component plasma (jellium)
in $\nu=2$ space dimensions.
The system is composed of $N$ mobile pointlike charges $e$, neutralized by
a uniform oppositely charged fixed background.
The total interaction potential $V(\vec{\bf r})$ satisfies for each of 
the particle coordinates the Poisson differential equation
\begin{equation} \label{2.9}
\enabla_j^2 V(\vec{\bf r}) = - 2\pi e^2 \sum_{k=1\atop(k\ne j)}^N 
\delta({\bf r}_j-{\bf r}_k) + 2\pi e^2 n ,
\quad j=1,2,\ldots,N .
\end{equation}
Here, the second term on the right-hand side (rhs) comes from 
the particle-background interaction and $n=N/\vert\Lambda\vert$ 
is the mean number density of the mobile charges. 
The summation over the particle index $j$ of the set of $N$ Poisson
equations (\ref{2.9}) results in
\begin{equation} \label{2.10}
\bfnabla^2 V(\vec{\bf r}) = - 2\pi e^2 \sum_{j,k=1\atop (j\ne k)}^N
\delta({\bf r}_j-{\bf r}_k) + 2\pi e^2 N n .
\end{equation}

The first term on the rhs of (\ref{2.10}), when weighted by the
classical Boltzmann factor 
$\propto \prod_{j<k} \vert {\bf r}_j-{\bf r}_k \vert^{\beta e^2}$
which vanishes at zero interparticle distance, does not give
any contribution to the Boltzmann density.
The application of the equality (\ref{2.7}) and the replacement
of $\bfnabla^2 V(\vec{\bf r})$ by the particle-background term
$2\pi N e^2 n$ in the relations (\ref{2.5}) and (\ref{2.6})
simplifies substantially the expansion formula for the quantum
Boltzmann density (\ref{2.8}):
\begin{eqnarray}
B_{\beta}(\vec{\bf r}) & = & \frac{1}{(\sqrt{2\pi}\lambda)^{\nu N}} \Bigg\{ 
{\rm e}^{-\beta V} \left( 1 - \frac{\lambda^2}{24} 2\pi\beta e^2 N n \right)
\prod_{j=1}^N \left( 1 - {\rm e}^{-2x_j^2/\lambda^2} \right) \nonumber \\
& & + \sum_{k=1}^N \prod_{j=1\atop (j\ne k)}^N 
\left( 1 - {\rm e}^{-2x_j^2/\lambda^2} \right) 
x_k {\rm e}^{-2x_k^2/\lambda^2}
\frac{\partial }{\partial x_k} {\rm e}^{-\beta V} \nonumber \\   
& & + \prod_{j=1}^N \left( 1 - {\rm e}^{-2x_j^2/\lambda^2} \right)
\frac{\lambda^2}{24} \bfnabla^2 {\rm e}^{-\beta V} \Bigg\} + o(\lambda^2) .
\label{2.11}
\end{eqnarray}

\section{Statistical quantities for the half-space geometry}
According to the standard formalism of statistical quantum mechanics,
the partition function of the $N$-particle fluid
(with ignored exchange effects) is given by the integration
of the quantum Boltzmann density over configuration space:
\begin{equation} \label{3.1}
Z_{\rm qu} = \frac{1}{N!} \int_{\Lambda} {\rm d}\vec{\bf r}\,
B_{\beta}(\vec{\bf r}) .
\end{equation}
The quantum average of a function $f(\vec{\bf r})$ is defined as follows
\begin{equation} \label{3.2}
\left\langle f \right\rangle_{\rm qu} = 
\frac{1}{Z_{\rm qu}} \frac{1}{N!} 
\int_{\Lambda} {\rm d}\vec{\bf r}\, 
B_{\beta}(\vec{\bf r}) f(\vec{\bf r}) .
\end{equation}
At the one-particle level, one introduces the particle number density
\begin{equation} \label{3.3}
n_{\rm qu}({\bf r}) = \Bigg\langle
\sum_{j=1}^N \delta({\bf r}-{\bf r}_j) \Bigg\rangle_{\rm qu} .
\end{equation}
At the two-particle level, one considers the two-body density
\begin{equation} \label{3.4}
n_{\rm qu}^{(2)}({\bf r},{\bf r}') = \Bigg\langle
\sum_{j,k=1\atop (j\ne k)}^N 
\delta({\bf r}-{\bf r}_j) \delta({\bf r}'-{\bf r}_k) \Bigg\rangle_{\rm qu} .
\end{equation}
It will be useful to consider also the truncated two-body density
\begin{equation} \label{3.5}
n_{\rm qu}^{(2){\rm T}}({\bf r}_1,{\bf r}_2) = 
n_{\rm qu}^{(2)}({\bf r}_1,{\bf r}_2)
- n_{\rm qu}({\bf r}_1) n_{\rm qu}({\bf r}_2) 
\end{equation}
vanishing at asymptotically large distances
$\vert {\bf r}_1-{\bf r}_2\vert\to\infty$.
The general multiparticle densities are defined in analogy with (\ref{3.4}),
i.e. the corresponding product of $\delta$-functions is summed out over 
all possible multiplets of different particles.

The classical partition function $Z$ and the classical average of a function
$f(\vec{\bf r})$ are defined as follows
\begin{eqnarray}
Z & = & \frac{1}{N!} \int_{\Lambda} 
\frac{{\rm d}\vec{\bf r}}{(\sqrt{2\pi}\lambda)^{\nu N}}\,
{\rm e}^{-\beta V(\vec{\bf r})} , \label{3.6} \\
\left\langle f \right\rangle & = &  \frac{1}{Z} \frac{1}{N!} 
\int_{\Lambda} \frac{{\rm d}\vec{\bf r}}{(\sqrt{2\pi}\lambda)^{\nu N}}\,
{\rm e}^{-\beta V(\vec{\bf r})} f(\vec{\bf r}) . \label{3.7}
\end{eqnarray}
The classical values of statistical quantities will be written 
without a subscript, like $n({\bf r})$, $n^{(2)}({\bf r},{\bf r}')$, etc.
In the calculations which follow, we shall need explicitly truncated forms 
of the classical three-body density
\begin{eqnarray} 
\fl n^{(3){\rm T}}({\bf r}_1,{\bf r}_2,{\bf r}_3) & = & 
n^{(3)}({\bf r}_1,{\bf r}_2,{\bf r}_3) - 
n^{(2){\rm T}}({\bf r}_1,{\bf r}_2) n({\bf r}_3) 
- n^{(2){\rm T}}({\bf r}_1,{\bf r}_3) n({\bf r}_2) \nonumber \\ & & 
- n^{(2){\rm T}}({\bf r}_2,{\bf r}_3) n({\bf r}_1) 
- n({\bf r}_1) n({\bf r}_2) n({\bf r}_3) \label{3.8}
\end{eqnarray}
and of the classical four-body density
\begin{eqnarray}
\fl n^{(4){\rm T}}({\bf r}_1,{\bf r}_2,{\bf r}_3,{\bf r}_4)
& = & n^{(4)}({\bf r}_1,{\bf r}_2,{\bf r}_3,{\bf r}_4) 
- n^{(3){\rm T}}({\bf r}_1,{\bf r}_2,{\bf r}_3) n({\bf r}_4) 
\nonumber \\ & & 
- n^{(3){\rm T}}({\bf r}_1,{\bf r}_2,{\bf r}_4) n({\bf r}_3) 
- n^{(3){\rm T}}({\bf r}_1,{\bf r}_3,{\bf r}_4) n({\bf r}_2) 
\nonumber \\ & & 
- n^{(3){\rm T}}({\bf r}_2,{\bf r}_3,{\bf r}_4) n({\bf r}_1) 
- n^{(2){\rm T}}({\bf r}_1,{\bf r}_2) n^{(2){\rm T}}({\bf r}_3,{\bf r}_4) 
\nonumber \\& &
- n^{(2){\rm T}}({\bf r}_1,{\bf r}_3) n^{(2){\rm T}}({\bf r}_2,{\bf r}_4)
- n^{(2){\rm T}}({\bf r}_1,{\bf r}_4) n^{(2){\rm T}}({\bf r}_2,{\bf r}_3) 
\nonumber \\ & &
- n^{(2){\rm T}}({\bf r}_1,{\bf r}_2) n({\bf r}_3) n({\bf r}_4)
- n^{(2){\rm T}}({\bf r}_1,{\bf r}_3) n({\bf r}_2) n({\bf r}_4)
\nonumber \\ & &
- n^{(2){\rm T}}({\bf r}_1,{\bf r}_4) n({\bf r}_2) n({\bf r}_3)
- n^{(2){\rm T}}({\bf r}_2,{\bf r}_3) n({\bf r}_1) n({\bf r}_4)
\nonumber \\ & &
- n^{(2){\rm T}}({\bf r}_2,{\bf r}_4) n({\bf r}_1) n({\bf r}_3)
- n^{(2){\rm T}}({\bf r}_3,{\bf r}_4) n({\bf r}_1) n({\bf r}_2)
\nonumber \\ & &
- n({\bf r}_1) n({\bf r}_2) n({\bf r}_3) n({\bf r}_4) . \label{3.9}
\end{eqnarray}

In what follows, we shall restrict ourselves to the model system
of the one-component plasma constrained to the two-dimensional 
half-space $\Lambda$. Now, the Cartesian coordinates of $\mathbf{r}$ become 
$(x,y)$, with the origin on the rectilinear plain hard wall, the $y$ axis
along the wall, and the system occupying the $x>0$ half-space $\Lambda$.

\subsection{Partition function}
Substituting the expansion of the Boltzmann density (\ref{2.11})
into the definition (\ref{3.1}) of the quantum partition function 
$Z_{\rm qu}$ and performing expansions of type (\ref{2.4}) for the
products of boundary factors, we obtain
\begin{eqnarray} 
\frac{Z_{\rm qu}}{Z} & = & 1 - \frac{\lambda^2}{24} 2\pi \beta e^2 N n
- \int_{\Lambda} {\rm d}{\bf r}_1\, {\rm e}^{-2x_1^2/\lambda^2} n({\bf r}_1)
\nonumber \\ & &
+ \frac{1}{2!} \int_{\Lambda}{\rm d}{\bf r}_1 \int_{\Lambda}{\rm d}{\bf r}_2\,
{\rm e}^{-2x_1^2/\lambda^2} {\rm e}^{-2x_2^2/\lambda^2}
n^{(2)}({\bf r}_1,{\bf r}_2) \nonumber \\ & &
+ \int_{\Lambda} {\rm d}{\bf r}_1\, {\rm e}^{-2x_1^2/\lambda^2}
x_1 \frac{\partial}{\partial x_1} n({\bf r}_1)
+ \frac{\lambda^2}{24} \int_{\Lambda} {\rm d}{\bf r}_1\, 
\frac{\partial^2}{\partial x_1^2} n({\bf r}_1) + o(\lambda^2) . \label{3.10}
\end{eqnarray} 
Here, we keep in mind that the integration of an exponential term
$\exp(-2x^2/\lambda^2)$ over $x$ produces one $\lambda$-factor.

\subsection{One-body density}
To calculate the quantum one-body density (\ref{3.3}), we take advantage
of the invariance of the Boltzmann density (\ref{2.11}) with respect to
permutations of the particle indices and write down
\begin{equation} \label{3.11}
n_{\rm qu}({\bf r}) = \frac{1}{Z_{\rm qu}} \frac{N}{N!}
\int_{\Lambda} {\rm d}\vec{\bf r}\, B_{\beta}(\vec{\bf r})
\delta({\bf r}-{\bf r}_1) .
\end{equation}
In each term of the Boltzmann density, we separate the ``reference'' 
${\bf r}_1$-dependent part, which is kept unchanged, and expand 
in analogy with (\ref{2.4}) the remaining part dependent on 
$({\bf r}_2,\ldots,{\bf r}_N)$ coordinates.
Like for instance,
\begin{eqnarray} 
\prod_{j=1}^N \left( 1 - {\rm e}^{-2x_j^2/\lambda^2} \right)
& = &  \left( 1 - {\rm e}^{-2x_1^2/\lambda^2} \right)
\Bigg\{ 1 - \sum_{j=2}^N {\rm e}^{-2x_j^2/\lambda^2} \nonumber \\
& & 
+ \frac{1}{2!} \sum_{j,k=2\atop (j\ne k)}^N {\rm e}^{-2x_j^2/\lambda^2} 
{\rm e}^{-2x_k^2/\lambda^2} + \cdots \Bigg\} .  \label{3.12}
\end{eqnarray}  
The quantum partition function $Z_{\rm qu}$ in the denominator on
the rhs of (\ref{3.11}) is substituted by the expansion (\ref{3.10})
and subsequently expanded in ``virtual'' $\lambda$ powers.
After simple but lengthy algebra, one obtains
\begin{eqnarray}
n_{\rm qu}({\bf r}_1) & = &
\left( 1 - {\rm e}^{-2x_1^2/\lambda^2} \right) \Bigg\{
n({\bf r}_1) - \int_{\Lambda} {\rm d}{\bf r}_2\, {\rm e}^{-2x_2^2/\lambda^2}
n^{(2){\rm T}}({\bf r}_1,{\bf r}_2) \nonumber \\  & &
+ \frac{1}{2!} \int_{\Lambda} {\rm d}{\bf r}_2 \int_{\Lambda} 
{\rm d}{\bf r}_3\, {\rm e}^{-2x_2^2/\lambda^2} {\rm e}^{-2x_3^2/\lambda^2}
n^{(3){\rm T}}({\bf r}_1,{\bf r}_2,{\bf r}_3) \nonumber \\ & &
+ \int_{\Lambda} {\rm d}{\bf r}_2\, {\rm e}^{-2x_2^2/\lambda^2}
x_2 \frac{\partial}{\partial x_2} n^{(2){\rm T}}({\bf r}_1,{\bf r}_2) 
\nonumber \\  & &
+ \frac{\lambda^2}{24} \left[ 
\frac{\partial^2}{\partial x_1^2} n({\bf r}_1)
+ \int_{\Lambda} {\rm d}{\bf r}_2\, \frac{\partial^2}{\partial x_2^2} 
n^{(2){\rm T}}({\bf r}_1,{\bf r}_2) \right] \Bigg\}
\nonumber \\  & &
+ {\rm e}^{-2x_1^2/\lambda^2} x_1 \frac{\partial}{\partial x_1}
\left[ n({\bf r}_1) - \int_{\Lambda} {\rm d}{\bf r}_2\, 
{\rm e}^{-2x_2^2/\lambda^2}
n^{(2){\rm T}}({\bf r}_1,{\bf r}_2) \right] . \label{3.13}
\end{eqnarray}

Let us assume that the classical averages under integrations in (\ref{3.13})
are analytic functions of the $x$-coordinate at the boundary $x=0$.
For instance, the classical density profile $n({\bf r})\equiv n(x)$
is supposed to exhibit the Taylor expansion
$n(x) = n(0) + n'(0)x + n''(0)x^2/2!+\cdots$.
Then, the integral
\begin{eqnarray}
\int_0^{\infty} {\rm d}x\, {\rm e}^{-2x^2/\lambda^2} n(x) & = &  
\lambda \int_0^{\infty} {\rm d}x'\, {\rm e}^{-2x'^2} n(\lambda x')
\nonumber \\ & = &
\lambda \frac{1}{2} \sqrt{\frac{\pi}{2}} n(0) + 
\lambda^2 \frac{1}{4} n'(0) + O(\lambda^3) . \label{3.14}
\end{eqnarray}  
Performing an analogous procedure in all integrals on the rhs of
(\ref{3.13}), we finally arrive at the expansion of the one-body 
density up to the $\lambda^2$ order:
\begin{eqnarray}
n_{\rm qu}({\bf r}_1) & = &
\left( 1 - {\rm e}^{-2x_1^2/\lambda^2} \right) \Bigg\{
n({\bf r}_1) - \lambda \sqrt{\frac{\pi}{8}} \int {\rm d}y_2\,
n^{(2){\rm T}}[{\bf r}_1,(0,y_2)] \nonumber \\ & &
+ \frac{\lambda^2}{2} \left( \frac{\pi}{8} \right)
\int {\rm d}y_2 \int {\rm d}y_3\,
n^{(3){\rm T}}[{\bf r}_1,(0,y_2),(0,y_3)]  \nonumber \\ & &
+ \frac{\lambda^2}{24} \left[ \frac{\partial^2}{\partial x_1^2} n({\bf r}_1)
- \int {\rm d}y_2\,
\frac{\partial n^{(2){\rm T}}[{\bf r}_1,(x_2,y_2)]}{\partial 
x_2} \Bigg\vert_{x_2=0} \right] \Bigg\}
\nonumber \\ & &
+ {\rm e}^{-2x_1^2/\lambda^2} x_1 \frac{\partial}{\partial x_1}
\left\{ n({\bf r}_1) - \lambda \sqrt{\frac{\pi}{8}}
\int {\rm d}y_2\, n^{(2){\rm T}}[{\bf r}_1,(0,y_2)]
\right\} . \label{3.15}
\end{eqnarray}

\subsection{Two-body density}
To calculate the $\lambda$-expansion of the quantum truncated two-body 
density given by relations (\ref{3.4}) and (\ref{3.5}), 
we proceed as in the previous subsection.
We first take advantage of the permutation invariance of the Boltzmann density 
(\ref{2.11}) to write down
\begin{equation} \label{3.16}
n_{\rm qu}^{(2)}({\bf r},{\bf r}') = \frac{1}{Z_{\rm qu}} \frac{N(N-1)}{N!}
\int_{\Lambda} {\rm d}\vec{\bf r}\, B_{\beta}(\vec{\bf r})
\delta({\bf r}-{\bf r}_1) \delta({\bf r}'-{\bf r}_2) .
\end{equation}
Then we separate the reference part in the Boltzmann density 
(\ref{2.11}) which depends on coordinates ${\bf r}_1, {\bf r}_2$ and
expand in analogy with (\ref{3.12}) the remaining part.
Finally, using the integration procedure of type (\ref{3.14}) in
all integrals we arrive at the expansion of the two-body density 
up to the $\lambda^2$ order:
\begin{eqnarray}
\fl n_{\rm qu}^{(2){\rm T}}({\bf r}_1,{\bf r}_2) 
& = & \left( 1 - {\rm e}^{-2x_1^2/\lambda^2} \right) 
\left( 1 - {\rm e}^{-2x_2^2/\lambda^2} \right) \Bigg\{
n^{(2){\rm T}}({\bf r}_1,{\bf r}_2) \nonumber \\ \fl & & 
- \lambda \sqrt{\frac{\pi}{8}} \int {\rm d}y_3\,
n^{(3){\rm T}}[{\bf r}_1,{\bf r}_2,(0,y_3)] \nonumber \\ \fl & & 
+ \frac{\lambda^2}{2} \left( \frac{\pi}{8} \right)
\int {\rm d}y_3 \int {\rm d}y_4\,
n^{(4){\rm T}}[{\bf r}_1,{\bf r}_2,(0,y_3),(0,y_4)]  
\nonumber \\ \fl & &
+ \frac{\lambda^2}{24} \left[ \left( \frac{\partial^2}{\partial x_1^2} + 
\frac{\partial^2}{\partial x_2^2} \right) 
n^{(2){\rm T}}({\bf r}_1,{\bf r}_2) - \int {\rm d}y_3\,
\frac{\partial n^{(3){\rm T}}[{\bf r}_1,{\bf r}_2,(x_3,y_3)]}
{\partial x_3} \Bigg\vert_{x_3=0} \right] \Bigg\}
\nonumber \\ \fl & &
+ \left( 1 - {\rm e}^{-2x_2^2/\lambda^2} \right) 
x_1 {\rm e}^{-2x_1^2/\lambda^2} \frac{\partial}{\partial x_1}
\Bigg\{ n^{(2){\rm T}}({\bf r}_1,{\bf r}_2) 
\nonumber \\ \fl & & 
\qquad - \lambda \sqrt{\frac{\pi}{8}} \int {\rm d}y_3\, 
n^{(3){\rm T}}[{\bf r}_1,{\bf r}_2,(0,y_3)] \Bigg\} 
\nonumber \\ \fl & &
+ \left( 1 - {\rm e}^{-2x_1^2/\lambda^2} \right) 
x_2 {\rm e}^{-2x_2^2/\lambda^2} \frac{\partial}{\partial x_2}
\Bigg\{ n^{(2){\rm T}}({\bf r}_1,{\bf r}_2)  
\nonumber \\ \fl & & 
\qquad - \lambda \sqrt{\frac{\pi}{8}} \int {\rm d}y_3\, 
n^{(3){\rm T}}[{\bf r}_1,{\bf r}_2,(0,y_3)] \Bigg\} . \label{3.17}
\end{eqnarray}

\section{Review of the sum rules}
The following sum rules [except the perfect-screening one (\ref{4.1})] apply
only to the one-component plasma (not to many-component ones); they rely on
the facts that, for the one-component plasma, the mass and charge fluctuations
are proportional to one another and the static resistivity vanishes. 
Although the sum rules can be written for a quantum one-component plasma 
in a $\nu$-dimensional half-space with any $\nu$, here we consider only the
case $\nu=2$. Some of the sum rules that we review were originally written
more generally for the time-displaced correlations, but here we consider only 
their static limits.

The perfect screening sum rule expresses that the charge cloud around a
particle of the system has a charge opposite to the charge of this particle. 
This sum rule has the same form as in the classical case:
\begin{equation}
\int\mathrm{d}\mathbf{r}_1 
n^{(2)\mathrm{T}}_{\mathrm{qu}}(\mathbf{r}_1,\mathbf{r}_2)=
-n_{\mathrm{qu}}(x_2).  \label{4.1}
\end{equation}
Although we have no doubt about the validity of (\ref{4.1}), for a
check of the calculations in section 3 and of their application to the present 
2D one-component plasma at $\beta e^2=2$, we shall verify 
(\ref{4.1}) at order $\lambda^2$ in section 6. 
 
Macroscopic arguments gave the asymptotic form of the quantum truncated
two-body density along the wall \cite{Janco2}
\begin{equation} \label{4.2}
n^{(2)\mathrm{T}}_{\mathrm{qu}}(\mathbf{r}_1,\mathbf{r}_2)
\mathop{\sim}_{|y_1-y_2|\rightarrow\infty}
\frac{f(x_1,x_2)}{(y_1-y_2)^2}, 
\end{equation}
(after perhaps an averaging on local oscillations in $y_1-y_2$) with the sum
rule 
\begin{equation} \label{4.3}
\fl \int_0^{\infty}\mathrm{d}x_1\int_0^{\infty}\mathrm{d}x_2 f(x_1,x_2) 
=-\frac{1}{4\pi^2 e^2}\left[2\hbar\omega_{\rm s}
\coth(\beta\hbar\omega_{\rm s}/2)
-\hbar\omega_{\rm p}\coth(\beta\hbar\omega_{\rm p}/2)\right] 
\end{equation}
(here $\beta e^2=2$), where the bulk and surface plasma frequencies, for two
dimensions and a plain hard wall, are, respectively,
\begin{equation} \label{4.4}
\omega_{\rm p}=\left(\frac{2\pi ne^2}{m}\right)^{1/2}, \quad
\omega_{\rm s}=\left(\frac{\pi ne^2}{m}\right)^{1/2}. 
\end{equation}
Any microscopic check of (\ref{4.3}) would be welcome. 
The expansion of the r.h.s. of (\ref{4.3}) in powers of $\hbar$ starts with
the classical value $-1/(4\pi^2)$ and the next term is of order $\hbar^4$. 
Therefore, at order $\lambda^2$, in section 7 we shall only be able
to check that there is no quantum correction.

Finally, another macroscopic argument gave the quantum form of 
the dipole sum rule \cite{JLM}
\begin{equation} \label{4.5}
\fl \int_0^{\infty}\mathrm{d}x_2\left[\int_0^{\infty}\mathrm{d}x_1 x_1
\int\mathrm{d}y_1 n^{(2)\mathrm{T}}_{\mathrm{qu}}(\mathbf{r}_1,\mathbf{r}_2)
+ x_2n_{\mathrm{qu}}(x_2)\right] 
=-\frac{\hbar\omega_{\mathrm{p}}}{4\pi e^2}
\coth\frac{\beta\hbar\omega_{\mathrm{p}}}{2} 
\end{equation}
(here $\beta e^2=2$). The quantity between square brackets in (\ref{4.5}) is
the dipole moment of the structure factor. Here too, a check would be welcome. 
Now, the expansion of the rhs of (\ref{4.5}) in powers of $\hbar$ starts 
with the classical value $-1/(4\pi)$ and the next term is 
$-\lambda^2/(12\pi a^2)$, where $a$ is the average interparticle 
distance defined by $n=1/(\pi a^2)$. 
In section 8 we can check this sum rule at order $\lambda^2$. 

\section{The classical densities}
We shall need some information about the classical densities of the
2D one-component plasma at $\beta e^2=2$. 
We express all lengths in units of the average interparticle distance $a$, 
thus $n=1/\pi$. 
The density profile is \cite{Janco}  
\begin{equation} \label{5.1}
n(x)=n\frac{2}{\sqrt{\pi}}\int_0^{\infty}\mathrm{d}t
\frac{\exp[-(t-x\sqrt{2})^2]}{1+\Phi(t)}, 
\end{equation}
where $\Phi(t)$ is the probability-integral function. 
The two-body truncated density is
\begin{equation} \label{5.2}
n^{(2)\mathrm{T}}(\mathbf{r}_1,\mathbf{r}_2)=-n^2\exp[-2(x_1^2+x_2^2)] 
k(\mathbf{r}_1,\mathbf{r}_2)k(\mathbf{r}_2,\mathbf{r}_1), 
\end{equation}
where
\begin{equation}
k(\mathbf{r}_1,\mathbf{r}_2)=\frac{2}{\sqrt{\pi}}\int_0^\infty\mathrm{d}t
\frac{\exp\left[-t^2+t(x_1+x_2)\sqrt{2}-\mathrm{i}t(y_1-y_2)\sqrt{2}\right]}
{1+\Phi(t)}.  \label{5.3}
\end{equation}
Higher-order $n$-body truncated densities contain a sum of product of $n$ 
factors $k$, the sum running on all oriented cycles built with 
$\mathbf{r}_1,\mathbf{r}_2,...,\mathbf{r}_n$ \cite{CJB}. One finds for the
three-body truncated density   
\begin{eqnarray}
n^{(3)\mathrm{T}}(\mathbf{r}_1,\mathbf{r}_2,\mathbf{r}_3) & = &
n^3\exp[-2(x_1^2+x_2^2+x_3^2)][k(\mathbf{r}_1,\mathbf{r}_2)
k(\mathbf{r}_2,\mathbf{r}_3)k(\mathbf{r}_3,\mathbf{r}_1) \nonumber \\
& & + k(\mathbf{r}_1,\mathbf{r}_3)k(\mathbf{r}_3,\mathbf{r}_2)
k(\mathbf{r}_2,\mathbf{r}_1)], \label{5.4}
\end{eqnarray}
and for the four-body truncated density
\begin{eqnarray}
\fl n^{(4)\mathrm{T}}(\mathbf{r}_1,\mathbf{r}_2,\mathbf{r}_3,\mathbf{r}_4) 
& = & -n^4\exp[-2(x_1^2+x_2^2+x_3^2+x_4^2)][k(\mathbf{r}_1,\mathbf{r}_2)
k(\mathbf{r}_2,\mathbf{r}_3)k(\mathbf{r}_3,\mathbf{r}_4)
k(\mathbf{r}_4,\mathbf{r}_1) \nonumber \\
& & + k(\mathbf{r}_1,\mathbf{r}_2)k(\mathbf{r}_2,\mathbf{r}_4)
k(\mathbf{r}_4,\mathbf{r}_3)k(\mathbf{r}_3,\mathbf{r}_1) \nonumber \\
& & + k(\mathbf{r}_1,\mathbf{r}_3)k(\mathbf{r}_3,\mathbf{r}_2)
k(\mathbf{r}_2,\mathbf{r}_4)k(\mathbf{r}_4,\mathbf{r}_1)] \nonumber \\
& & + \mbox{complex conjugate}. \label{5.5}
\end{eqnarray}

We shall also need some integrals of these densities. 
In (\ref{5.2}), the product of the $k$ functions contains 
$\int_0^{\infty}\mathrm{d}t\exp(-\mathrm{i}ty_1\sqrt{2})
\int_0^{\infty}\mathrm{d}t'\exp(\mathrm{i}t'y_1\sqrt{2})$, 
therefore the integral of this quantity on $y_1$ is 
$\pi\sqrt{2}\,\delta(t-t')$. One obtains
\begin{equation} \label{5.6}
\fl \int\mathrm{d}y_1n^{(2)\mathrm{T}}(\mathbf{r}_1,\mathbf{r}_2) =
- n^2 4\sqrt{2}\exp[-2(x_1^2+x_2^2)] \int_0^{\infty}\mathrm{d}t
\frac{\exp\left[-2t^2+t(x_1+x_2)2\sqrt{2}\right]}{[1+\Phi(t)]^2}. 
\end{equation}
By the same method, one finds
\begin{eqnarray}
\fl \int\mathrm{d}y_3n^{(3)\mathrm{T}}(\mathbf{r}_1,\mathbf{r}_2,\mathbf{r}_3)
& = & n^3\frac{8\sqrt{2}}{\sqrt{\pi}}\exp[-2(x_1^2+x_2^2+x_3^2)] \nonumber \\
& &\times\int_0^{\infty}\mathrm{d}t
\frac{\exp[-t^2+t(x_1+x_2)\sqrt{2}-\mathrm{i}t(y_1-y_2)\sqrt{2}]}
{1+\Phi(t)} \nonumber \\
& &\times\int_0^{\infty}\mathrm{d}t'
\frac{\exp[-2t'^2+t'(x_1+x_2+2x_3)\sqrt{2}
-\mathrm{i}t'(y_2-y_1)\sqrt{2}]}{[1+\Phi(t')]^2} \nonumber \\
& & + \mbox{complex conjugate}, \label{5.7}
\end{eqnarray}
\begin{eqnarray}
\fl \int\mathrm{d}y_1\int\mathrm{d}y_3n^{(3)\mathrm{T}}
(\mathbf{r}_1,\mathbf{r}_2,\mathbf{r}_3) & = &
n^3 32\sqrt{\pi}\exp[-2(x_1^2+x_2^2+x_3^2)] \nonumber \\
& & \times\int_0^{\infty}\mathrm{d}t
\frac{\exp[-3t^2+t(x_1+x_2+x_3)2\sqrt{2}]}{[1+\Phi(t)]^3},
\label{5.8}
\end{eqnarray}
\begin{eqnarray}
\fl \int\mathrm{d}y_3\int\mathrm{d}y_4n^{(4)\mathrm{T}}
[\mathbf{r}_1,\mathbf{r}_2,(0,y_3),(0,y_4)] =
- n^4 32\exp[-2(x_1^2+x_2^2)] \nonumber \\
\times\left\{2\int_0^{\infty}\mathrm{d}t
\frac{\exp[-t^2+t(x_1+x_2)\sqrt{2}-\mathrm{i}t(y_1-y_2)\sqrt{2}]}{1+\Phi(t)}
\right. \nonumber \\
\times\int_0^{\infty}\mathrm{d}t'\frac{\exp[-3t'^2+t'(x_1+x_2)\sqrt{2}
-\mathrm{i}t'(y_2-y_1)\sqrt{2}]}{[1+\Phi(t')]^3} \nonumber \\
+\int_0^{\infty}\mathrm{d}t\frac{\exp[-2t^2+t(x_1+x_2)\sqrt{2}
-\mathrm{i}t(y_1-y_2)\sqrt{2}]}{[1+\Phi(t)]^2} \nonumber \\
\left.\times\int_0^{\infty}\mathrm{d}t'\frac{\exp[-2t'^2+t'(x_1+x_2)\sqrt{2}
-\mathrm{i}t'(y_2-y_1)\sqrt{2}]}{[1+\Phi(t')]^2}\right\} \nonumber \\
+  \mbox{complex conjugate}, \label{5.9}
\end{eqnarray}
\begin{eqnarray}
\fl \int\mathrm{d}y_1\int\mathrm{d}y_3\int\mathrm{d}y_4n^{(4)\mathrm{T}}
[\mathbf{r}_1,\mathbf{r}_2,(0,y_3),(0,y_4)] = 
- n^4 192\pi\sqrt{2}\exp[-2(x_1^2+x_2^2)] \nonumber \\
\times \int_0^{\infty}\mathrm{d}t\frac{\exp[-4t^2+t(x_1+x_2)2\sqrt{2}]}
{[1+\Phi(t)]^4}. \label{5.10}
\end{eqnarray}

\section{Perfect screening}
Omitting some terms which do not contribute to the sum rules at order
$\lambda^2$, and using the results in section 5, we find from (3.17)
\begin{eqnarray}
\fl \int\mathrm{d}y_1 
n_{\mathrm{qu}}^{(2)\mathrm{T}}(\mathbf{r}_1,\mathbf{r}_2)
&&=[1-\exp(-2x_1^2/\lambda^2)][1-\exp(-2x_2^2/\lambda^2)] \nonumber \\
&&\times\exp[-2(x_1^2+x_2^2)]\left\{-n^2 4\sqrt{2}\int_0^{\infty}
\mathrm{d}t\frac{\exp[-2t^2+t(x_1+x_2)2\sqrt{2}]}{[1+\Phi(t)]^2}\right. 
\nonumber \\
&&-\lambda n^3 8\sqrt{2}\,\pi\int_0^{\infty}\mathrm{d}t
\frac{\exp[-3t^2+t(x_1+x_2)2\sqrt{2}]}{[1+\Phi(t)]^3} \nonumber \\
&&-\lambda^2 n^4 12\sqrt{2}\,\pi^2\int_0^{\infty}\mathrm{d}t
\frac{\exp[-4t^2+t(x_1+x_2)2\sqrt{2}]}{[1+\Phi(t)]^4} \nonumber \\
&&\left.-\lambda^2 n^3\frac{8}{3}\sqrt{2\pi}\int_0^{\infty}\mathrm{d}t\,t
\frac{\exp[-3t^2+t(x_1+x_2)2\sqrt{2}]}{[1+\Phi(t)]^3}\right\} \nonumber \\
&&+\left\{-\lambda^2 n^2\frac{\sqrt{2}}{6}\left(\frac{\partial^2}
{\partial x_1^2}+\frac{\partial^2}{\partial x_2^2}\right)
-n^2 4\sqrt{2}\exp(-2x_1^2/\lambda^2)x_1\frac{\partial}{\partial x_1}\right.
\nonumber \\
&&\left.-n^2 4\sqrt{2}\exp(-2x_2^2/\lambda^2)x_2\frac{\partial}{\partial x_2}
\right\} \nonumber \\
&&\times\left\{\exp[-2(x_1^2+x_2^2)]
\int_0^{\infty}\mathrm{d}t\frac{\exp[-2t^2+t(x_1+x_2)2\sqrt{2}]}{[1+\Phi(t)]^2}
\right\}+\cdots\; . \label{6.1}
\end{eqnarray}
On the other hand, we find from (3.15)
\begin{eqnarray}
\fl n_{\mathrm{qu}}(x_2) & = & 
[1-\exp(-2x_2^2/\lambda^2)]\left\{\frac{2}{\sqrt{\pi}} n\exp(-2x_2^2)
\int_0^{\infty}\mathrm{d}t\frac{\exp(-t^2+tx_2 2\sqrt{2})}{1+\Phi(t)}
\right.\nonumber \\
\fl & &+\lambda n^2 2\sqrt{\pi}\exp(-2x_2^2) 
\int_0^{\infty}\mathrm{d}t\frac{\exp(-2t^2+tx_2 2\sqrt{2})}{[1+\Phi(t)]^2}
\nonumber \\
\fl & &+\lambda^2 n^3 2\pi^{3/2}\exp(-2x_2^2) 
\int_0^{\infty}\mathrm{d}t\frac{\exp(-3t^2+tx_2 2\sqrt{2})}{[1+\Phi(t)]^3}
\nonumber \\
\fl & &+\lambda^2 n\frac{1}{12\sqrt{\pi}}\frac{\partial^2}{\partial x_2^2}
\left[\exp(-2x_2^2)
\int_0^{\infty}\mathrm{d}t\frac{\exp(-t^2+tx_2 2\sqrt{2})}{1+\Phi(t)}\right]
\nonumber \\
\fl & &\left.+\lambda^2 n^2\frac{2}{3}\exp(-2x_2^2)
\int_0^{\infty}\mathrm{d}t\,t\frac{\exp(-2t^2+tx_2 2\sqrt{2})}{[1+\Phi(t)]^2}
\right\} \nonumber \\
\fl & &+\exp(-2x_2^2/\lambda^2)x_2\frac{\partial}{\partial x_2}
\left\{n\frac{2}{\sqrt{\pi}}\exp(-2x_2^2) \int_0^{\infty}
\mathrm{d}t\frac{\exp(-t^2+tx_2 2\sqrt{2})}{1+\Phi(t)} \right. \nonumber \\
\fl & &\left.+\lambda n^2 2\sqrt{\pi}\exp(-2x_2^2)\int_0^{\infty}\mathrm{d}t
\frac{\exp(-2t^2+tx_2 2\sqrt{2})}{[1+\Phi(t)]^2}\right\}+\cdots\; . \label{6.2}
\end{eqnarray}
Using \cite{GR}
\begin{equation}
\int_0^{\infty}\mathrm{d}x\exp(-2x^2+tx2\sqrt{2})=\sqrt{\frac{\pi}{8}}
\exp(t^2)[1+\Phi(t)], \label{6.3} 
\end{equation}
we can compute the integral on $x_1$ of (\ref{6.1}) at order $\lambda^2$
and check that it is equal to the opposite of (\ref{6.2}).

The sum rule (\ref{4.1}) is verified.

\section{The sum rule about an asymptotic form}
For obtaining the asymptotic forms of the classical $n$-body densities as
$|y_1-y_2|\rightarrow\infty$, one uses the integration per partes for the
Fourier transform of a function $F(t)$
\begin{equation} \label{7.1}
\int_0^{\infty}\mathrm{d}tF(t)\exp[-\mathrm{i}ty\sqrt{2}]\sim
\frac{F(0)}{\mathrm{i}y\sqrt{2}}. 
\end{equation}
The different classical $n$-body truncated densities or their integrals which
appear in (3.17) are all found to have an asymptotic form $\propto
\exp[-2(x_1^2+x_2^2)]/(y_1-y_2)^2$.
This gives the asymptotic form (\ref{4.2}), where (some terms which do not
contribute to the sum rule at order $\lambda^2$ have not been kept)
\begin{eqnarray}
f(x_1,x_2) & = & [1-\exp(-2x_1^2/\lambda^2)][1-\exp(-2x_2^2/\lambda^2)]
\nonumber \\ & & \times \exp[-2(x_1^2+x_2^2)]
\left( -n^2\frac{2}{\pi}-\lambda n^3 4 -\lambda^2 n^4 6\pi \right) 
+ \cdots\; . \label{7.2}
\end{eqnarray}
Now, we note from equation (\ref{3.14}) that, at order $\lambda^2$, 
an integral of the form 
$\int_0^{\infty}\mathrm{d}x\exp(-2x^2/\lambda^2)F(x)$ becomes
$\lambda\sqrt{\pi/8}F(0)+\lambda^2(1/4)F'(0)$. 
Then, from (\ref{7.2}), taking into account that in our units $n=1/\pi$,  
\begin{eqnarray}
\int_0^{\infty}\mathrm{d}x_1\int_0^{\infty}\mathrm{d}x_2 f(x_1,x_2)
& = &
-\frac{1}{4\pi^2} +
\lambda \left( \frac{1}{4\pi^2}+\frac{1}{4\pi^2}-\frac{1}{2\pi^2} \right)
\nonumber \\ 
& & +\lambda^2 \left( -\frac{1}{4\pi^2}+\frac{1}{2\pi^2}+\frac{1}{2\pi^2}
-\frac{3}{4\pi^2} \right) + o(\lambda^2) \nonumber \\
& = & -\frac{1}{4\pi^2}+o(\lambda^2) \label{7.3}
\end{eqnarray}
where $\lambda$ is in units of $a$. 

At order $\lambda^2$, there are no quantum corrections, in agreement with
section 4. 

\section{Dipole sum rule}
Omitting some terms which do not contribute to the sum rule at order
$\lambda^2$, using \cite{GR}
\begin{equation}
\int_0^{\infty}\mathrm{d}x\,x\exp(-2x^2+2\sqrt{2}\,tx)=
\frac{1}{4}+\frac{\sqrt{\pi}}{4}t\exp(t^2)[1+\Phi(t)], \label{8.1}
\end{equation}
and noting that, at order $\lambda^2$, an integral of the form 
$\int_0^{\infty}\mathrm{d}x\,x\exp(-2x^2/\lambda^2)F(x)$ becomes 
$(1/4)\lambda^2 F(0)$, one finds from (\ref{6.1}), contributing to the sum
rule at order $\lambda^2$,
\begin{eqnarray}
\fl \int_0^{\infty}\mathrm{d}x_1\,x_1\int\mathrm{d}y_1 
n^{(2)\mathrm{T}}_{\mathrm{qu}}(\mathbf{r}_1,\mathbf{r}_2) = 
[1-\exp(-2x_2^2/\lambda^2)]\exp(-2x_2^2) \nonumber \\
\times\left\{-n^2\sqrt{2}\int_0^{\infty}\mathrm{d}t
\left[\frac{\exp(-2t^2+tx_2 2\sqrt{2})}{[1+\Phi(t)]^2}
+\sqrt{\pi}\,t\frac{\exp(-t^2+tx_2 2\sqrt{2})}{1+\Phi(t)}\right] \right. 
\nonumber \\
+\lambda^2 n^2\sqrt{2}\int_0^{\infty}\mathrm{d}t
\frac{\exp(-2t^2+tx_2 2\sqrt{2})}{[1+\Phi(t)]^2} \nonumber \\
-\lambda n^3 2\sqrt{2}\pi
\int_0^{\infty}\mathrm{d}t\left[\frac{\exp(-3t^2+tx_2 2\sqrt{2})}
{[1+\Phi(t)]^3}+\sqrt{\pi}\,t\frac{\exp(-2t^2+tx_2 2\sqrt{2})}
{[1+\Phi(t)]^2}\right] \nonumber \\
-\lambda^2 n^4 3\sqrt{2}\pi^2 \int_0^{\infty}\mathrm{d}t 
\left[\frac{\exp(-4t^2+tx_2 2\sqrt{2})}{[1+\Phi(t)]^4}+
\sqrt{\pi}\,t\frac{\exp(-3t^2+tx_2 2\sqrt{2})}{[1+\Phi(t)]^3}\right] 
\nonumber \\
\left.-\lambda^2 n^3 \frac{2}{3}\sqrt{2\pi}\int_0^{\infty}\mathrm{d}t\,t
\left[\frac{\exp(-3t^2+tx_2 2\sqrt{2})}{[1+\Phi(t)]^3}+
\sqrt{\pi}\,t\frac{\exp(-2t^2+tx_2 2\sqrt{2})}{[1+\Phi(t)]^2}\right]\right\}
\nonumber \\
-\lambda^2 n^2 \frac{\sqrt{2}}{6}\exp(-2x_2^2)\int_0^{\infty}\mathrm{d}t
\frac{\exp(-2t^2+tx_2 2\sqrt{2})}{[1+\Phi(t)]^2} \nonumber \\
+\left\{-\lambda^2 n^2\frac{\sqrt{2}}{24}\frac{\partial^2}{\partial x_2^2}
-n^2\sqrt{2}\exp(-2x_2^2/\lambda^2)x_2\frac{\partial}{\partial x_2}\right\}
\nonumber \\
\times\left\{\exp(-2x_2^2)\int_0^{\infty}\mathrm{d}t
\left[\frac{\exp(-2t^2+tx_2 2\sqrt{2})}{[1+\Phi(t)]^2}\right.\right.
\nonumber \\
\left.\left.+\sqrt{\pi}\,t\frac{\exp(-t^2+tx_2 2\sqrt{2})}{1+\Phi(t)}
\right]\right\}+\cdots\; . \label{8.2}
\end{eqnarray}
On the other hand, $n_{\mathrm{qu}}(x_2)$ is given by (\ref{6.2}).

The classical part of the sum rule comes from the second line of
(\ref{8.2}) and the first line of (\ref{6.2}). 
Since $\exp(-t^2)/[1+\Phi(t)]^2=-(\sqrt{\pi}/2)(\mathrm{d}/\mathrm{d}t)
[1+\Phi(t)]^{-1}$, an integration per partes gives, with $n=1/\pi$,
\begin{equation}
\int_0^{\infty}\mathrm{d}x_1x_1\int\mathrm{d}y_1 
n^{(2)\mathrm{T}}(\mathbf{r}_1,\mathbf{r}_2)+x_2 n(x_2)=
-\frac{1}{\sqrt{2}\,\pi^{3/2}}\exp(-2x_2^2). \label{8.3}
\end{equation}
Integrating on $x_2$ gives the classical result $-1/(4\pi)$.

Using (\ref{6.3}) and (\ref{8.1}), it is straightforward to integrate on $x_2$
most of the terms of (\ref{8.2}) and $x_2$ times (\ref{6.2}). 
The term in (\ref{8.2}), line $-3$, of the form 
$\partial^2 F(x_2)/\partial x_2^2$ requires some care for evaluating 
its integral on $x_2$, which is
$\partial F(x_2)/\partial x_2 \left|_0^{\infty}\right.$. 
The point is that $F(x_2)$ has a term which is not zero at infinity. 
Indeed
\begin{eqnarray}
\fl \exp(-2x_2^2)\int_0^{\infty}\mathrm{d}t\,t\frac{\exp(-t^2+tx_2 2\sqrt{2})}
{1+\Phi(t)} & = & \int_0^{\infty}\mathrm{d}t\,t\frac{\exp[-(t-x_2\sqrt{2})^2]}
{1+\Phi(t)} \nonumber \\ & &
\mathop{\sim}_{x_2\rightarrow\infty}x_2\sqrt{2}
\int_{-\infty}^{\infty}\mathrm{d}t' \frac{\exp(-t'^2)}{2} 
= \sqrt{\pi/2}\,x_2  \label{8.4}
\end{eqnarray}
where $t'=t-x_2\sqrt{2}$, and the derivative with respect to $x_2$ 
of (\ref{8.4}) is $\sqrt{\pi/2}$ at infinity. 
The contribution of this derivative at infinity to the sum rule is found 
to be $-\lambda^2/(24\pi)$. 
Similarly,  the term in (\ref{6.2}), line $-2$, gives a contribution of 
the form $x_2[\partial^2 G(x_2)/\partial x_2^2]$; when integrated per partes 
on $x_2$, in particular it gives a term $-G(\infty)$. 
Since 
\begin{equation} \label{8.5}
\fl G(x_2)=\int_0^{\infty}\mathrm{d}t\frac{\exp[-(t-x_2\sqrt{2})^2]}{1+\Phi(t)}
\mathop{\sim}_{x_2\rightarrow\infty} 
\int_{-\infty}^{\infty}\mathrm{d}t'\frac{\exp(-t'^2)}{2}=\sqrt{\pi}/2,
\end{equation}
the corresponding contribution to the sum rule is $-\lambda^2/(24\pi)$. 
All other quantum contributions to the sum rule are found to cancel each
other, at order $\lambda^2$. 
Finally,
\begin{equation} \label{8.6}
\fl \int_0^{\infty}\mathrm{d}x_2\left[\int_0^{\infty}\mathrm{d}x_1x_1
\int\mathrm{d}y_1 n^{(2)\mathrm{T}}_{\mathrm{qu}}(\mathbf{r}_1,\mathbf{r}_2)
+x_2 n_{\mathrm{qu}}(x_2)\right]=-\frac{1}{4\pi}-\frac{\lambda^2}{12\pi}
+o(\lambda^2) . 
\end{equation}

Since $\lambda$ is in units of $a$, the sum rule (\ref{4.5}) does
have a quantum correction $-\lambda^2/(12\pi a^2)$ at order $\lambda^2$.

\section{Conclusion}
That the perfect screening sum rule (\ref{4.1}) is satisfied is no surprise.
This is rather a check of our calculations. The other sum rules are less
straightforward. In their heuristic macroscopic derivations~\cite{Janco2,JLM},
an essential feature of the quantum systems, that the $n$-body densities have
to vanish on the wall, was not explicitly taken into account. Thus a check
that they are indeed correct is welcome.  

The case of the asymptotic-form sum rule (\ref{4.3}) is not entirely
satisfactory. Although we have checked it at order $\lambda^2$, at this order
we could only verify that there is no quantum correction, in agreement with
the expansion of the rhs of (\ref{4.3}). It would have been more satisfactory
to verify the quantum correction of order $\lambda^4$ of this rhs. 
Unfortunately, this would involve pushing the Wigner-Kirkwood expansion
to that order, which would be straightforward but so tedious that we cannot
hope for its feasibility in a reasonable time.

The dipole sum rule (\ref{4.5}) is more tractable, and we have indeed checked
a finite quantum correction at order $\lambda^2$. Moreover, another derivation 
of this sum rule is feasible. Indeed, the generalization of the classical
Stillinger-Lovett sum rule~\cite{SL} for a bulk quantum one-component plasma
has been microscopically done~\cite{MO}; adapting the result to two dimensions
gives, with $\mathbf{r}=\mathbf{r}_1-\mathbf{r}_2$,
\begin{equation}
\int\mathrm{d}\mathbf{r}\,\mathbf{r}^2 n_{\mathrm{qu,\,bulk}}^{(2)\mathrm{T}}
(\mathbf{r}_1,\mathbf{r}_2)=-\frac{1}{\pi e^2}\hbar\omega_{\mathrm{p}}
\coth\frac{\beta\hbar\omega_{\mathrm{p}}}{2}. \label{9.1}
\end{equation}
Then, we can use the same kind of argument as the one in pages 965 and 966 of
ref.~\cite{JLM} for deriving the dipole sum rule (\ref{4.5}).

\ack
We gratefully acknowledge the support received from the European Science 
Foundation (ESF ``Methods of Integrable Systems, Geometry,  
Applied Mathematics'') and from the VEGA grant 2/6071/2006 of 
the Slovak Grant Agency.

\end{document}